\documentclass[conference]{IEEEtran}
\IEEEoverridecommandlockouts

\usepackage{array}
\newcolumntype{P}[1]{>{\centering\arraybackslash}m{#1}}
\usepackage{multirow}

\usepackage{cite}
\usepackage{amsmath,amssymb,amsfonts}
\usepackage{algorithm}
\usepackage{algpseudocode}
\usepackage{graphicx}
\usepackage{textcomp}
\usepackage{xcolor}
\usepackage[flushleft]{threeparttable}
\usepackage{enumitem}
\usepackage{{booktabs}}
\usepackage[bottom]{footmisc}

\usepackage{url}
\usepackage{hyperref}

\usepackage{tikz}
\definecolor{applegreen}{rgb}{0.55,0.71,0.0}
\newcommand{\greencircled}[1]{\tikz[baseline=(char.base)]{\node[shape=circle,draw=none,fill=applegreen,inner sep=1.2pt] (char) {\textcolor{white}{\small\bfseries #1}};}}

\newcommand{\orangecircled}[1]{\tikz[baseline=(char.base)]{\node[shape=circle,draw=none,fill=orange,inner sep=1.2pt] (char) {\textcolor{white}{\small\bfseries #1}};}}

\newcommand{\redcircled}[1]{\tikz[baseline=(char.base)]{\node[shape=circle,draw=none,fill=red,inner sep=1.2pt] (char) {\textcolor{white}{\small\bfseries #1}};}}

\def\BibTeX{{\rm B\kern-.05em{\sc i\kern-.025em b}\kern-.08em
    T\kern-.1667em\lower.7ex\hbox{E}\kern-.125emX}}
\begin{document}

\title{WeldMon: A Cost-effective Ultrasonic Welding Machine Condition Monitoring System}

\author{
\IEEEauthorblockN{Beitong Tian\textsuperscript{1}, Kuan-Chieh Lu\textsuperscript{2}, Ahmadreza Eslaminia\textsuperscript{2}, Yaohui Wang\textsuperscript{1}, Chenhui Shao\textsuperscript{2}, Klara Nahrstedt\textsuperscript{1}}
\IEEEauthorblockA{\textit{\textsuperscript{1}Coordinated Science Laboratory, \textsuperscript{2}Department of Mechanical Science and Engineering} \\
\textit{University of Illinois at Urbana-Champaign}, Champaign, USA \\
\textit{\{beitong2,kclu3,ae15,yaohuiw2,chshao,klara\}@illinois.edu}}
}

\maketitle

\begin{abstract}
Ultrasonic welding machines play a critical role in the lithium battery industry, facilitating the bonding of batteries with conductors. Ensuring high-quality welding is vital, making tool condition monitoring systems essential for early-stage quality control. However, existing monitoring methods face challenges in cost, downtime, and adaptability. In this paper, we present \textbf{WeldMon}, an affordable ultrasonic welding machine condition monitoring system that utilizes a custom data acquisition system and a data analysis pipeline designed for real-time analysis. Our classification algorithm combines auto-generated features and hand-crafted features, achieving superior cross-validation accuracy (95.8\% on average over all testing tasks) compared to the state-of-the-art method (92.5\%) in condition classification tasks. Our data augmentation approach alleviates the concept drift problem, enhancing tool condition classification accuracy by 8.3\%. All algorithms run locally, requiring only 385 milliseconds to process data for each welding cycle. We deploy WeldMon and a commercial system on an actual ultrasonic welding machine, performing a comprehensive comparison. Our findings highlight the potential for developing cost-effective, high-performance, and reliable tool condition monitoring systems.
\end{abstract}

\begin{IEEEkeywords}
Tool Condition Monitoring, Data Acquisition System, Concept Drift, Edge Computing
\end{IEEEkeywords}

\section{Introduction}
\label{sec:introduction}

Ultrasonic welding machines (UWM) play a critical role in the lithium battery industry, as they are used to bond batteries with conductors through a process known as the welding cycle. The welding cycle involves several steps, including setup, preparation, welding, cooling, and post-welding operations. The welding phase, highlighted in the green box in Fig. \ref{fig:weldmon_architecture}\textbf{(a)}, is the most crucial step. During this phase, the horn presses two conductor pieces onto the anvil and strikes them with a frequency of 20kHz for about one second, bonding the conductors.
High-quality welding is essential. Otherwise, the entire battery set may become unusable. Welding quality is influenced by factors such as tool condition, materials used, conductor surface condition, welding machine parameters, and the human factor. Among these factors, tool condition is the most critical, making UWM tool condition monitoring systems vital for early-stage quality control.

Various UWM tool condition monitoring methods have been extensively researched in academia. For instance, 3D scanned images of the horn and anvil can be used to inspect tool conditions directly. This approach, known as an offline method, is the most straightforward way to assess wear but requires specialized camera equipment and taking the horn and anvil out of the welding machine, causing significant cost and downtime.
Another research focuses on online methods that indirectly monitor UWM conditions by gathering and analyzing information from multiple sensors attached to or near the UWM. For example, \cite{nazir2021online} employs acoustic emission (AE), power, LVDT (linear velocity displacement transducer), and microphone sensors to collect multi-modal data during the welding phase. This data is then used in a supervised learning fashion to classify tool, material, and surface conditions or their combinations. While this method offers good classification accuracy and minimal downtime, it relies on commercial systems, specifically expensive industrial-grade data acquisition systems (costing over \$3,500) and hand-crafted features, which may not be optimal. Moreover, the method does not account for the influence of concept drift on accuracy. Concept drift describes a practical issue in data-driven methods, where testing data deviates from training data due to changes in time or environment, resulting in decreased testing phase accuracy. A naïve solution involves retraining the model for the new context. However, the substantial labor and time required for data collection and labeling,  as well as the difficulty in determining when to retrain, render this approach impractical.

In this paper, we introduce \textbf{WeldMon}, a cost-effective ultrasonic \textbf{Weld}ing machine condition \textbf{Mon}itoring system that can run on the edge device in a standalone way and provide robust and accurate monitoring results in real-time. We ingeniously utilize off-the-shelf sensors, audio ADCs (analog to digital converters), and single-board computers to create a data acquisition system that is significantly more affordable while meeting the specifications required for UWM condition monitoring.
We convert multi-modal sensory data collected during each welding phase into a multi-channel image and employ neural network models used for computer vision tasks for feature extraction. The auto-generated features, together with hand-crafted features, are then used for tool condition classification. 
We design three classification tasks to evaluate our method under realistic scenarios and compare it with various algorithms. Additionally, we assess the differences between WeldMon and a baseline commercial system, thoroughly testing the optimal sensor combinations and the necessity for data augmentation.

In summary, our work has the following contribution: 
\begin{enumerate}[leftmargin=*,label= ({\textbf{\arabic*}})]
\item We build a cost-effective data acquisition system for UWM condition monitoring. The system's price (as low as \$120) is more than 20 times less than the widely used system in academics ($>$ \$3500) with only a 1\% accuracy drop when used on the tool condition classification task. 
\item We propose a novel neural network-based data processing and augmentation pipeline for classifying UWM conditions. Our approach outperforms the previous state-of-the-art method in most test cases, offers greater robustness to context changes, and enables real-time classification on edge devices.
\item We implement and deploy WeldMon to a welding machine located in a research lab and comprehensively evaluate our system with real-world collected data.
\end{enumerate}

\section{Background \& Scope}
\label{sec:background_scope}

\subsection{Potential Factors Causing Concept Drift}
In addition to UWM tool condition (new/worn), various factors can influence the readings of sensors monitoring the UWM. In this subsection, we will discuss these factors, and their impact on sensor readings, and provide examples to illustrate the changes.

\noindent\textbf{Welding Parameters:}
Different welding tasks require specific sets of parameters, such as welding time, amplitude (measured in micrometers), and pressure (measured in Psi). While welding time and pressure may not significantly impact sensor readings, the amplitude can directly affect the readings of all sensors. Different UWM models will cause substantial changes in sensor readings and sensor deployment locations. However, for different UWMs of the same model, based on our experience, sensor readings will barely change as long as sensors are deployed in similar locations.

\noindent\textbf{Workpiece Characteristics:}
The material (copper/aluminum), shape, and dimensions of the workpieces used in each welding process can vary. However, our experience shows that sensor readings are almost identical regardless of these variations. One critical factor affecting sensor readings is the cleanliness of the workpiece surface. We simulated contaminated surface conditions by dropping cutting fluid on the welding spot and collected accelerometer data under both clean and contaminated conditions. We then visualized the results using a spectrogram, as shown in Fig. \ref{fig:all_spectrograms}. We observed a significant change in the sensor reading pattern when the surface was contaminated, regardless of tool condition (new or worn). It's important to note that these new patterns may not have been seen in the model development phase, potentially leading to incorrect decisions made by the model. In addition to that, in some cases, when the workpiece is contaminated, the spectrograms of new and worn conditions are barely distinguishable. 

\subsection{Properties of Sensors and ADCs}
Sensors and ADCs are key components in the UWM condition monitoring system. This subsection provides an overview of their properties, as well as how to leverage them for reducing costs.

Sensors are responsible for converting measured physical quantities into corresponding electrical signals, such as voltage. They can be classified into two categories: digital and analog sensors. Digital sensors output digital data, which is converted by an internal ADC, and have limited bandwidth due to the internal ADCs they use compared to analog sensors. Analog sensors, on the other hand, output a scaled analog voltage directly correlated to the measured physical quantity. Manufacturers typically provide a maximum sensor bandwidth recommendation, as oversampling may not yield useful information due to factors such as noise, small amplitude, or other unknown issues caused by oversampling. However, it is possible to modify some specific sensors' application circuitry and oversample these sensors, thereby extracting valuable information from higher frequency ranges.

ADCs are essential components in DAQ devices, responsible for sampling voltage signals from analog sensors. ADCs used in DAQ devices should accurately sample multiple voltages simultaneously at very high frequencies (up to several hundred thousand hertz) with minimal noise. A general-purpose ADC is commonly the preferred choice for constructing DAQ devices due to its applicability across various scenarios. In addition to general-purpose ADCs, there are specialized ADCs, such as audio ADCs. Designed explicitly for sampling output voltage from microphones, audio ADCs can easily sample at 100 kHz at a lower cost. Although audio ADCs have certain limitations compared to general-purpose ADCs, these limitations do not adversely affect our applications. 

\subsection{Scope of WeldMon}
In WeldMon, our goal is to develop a UWM condition monitoring system capable of determining the current tool condition (new or worn) after each welding process. We employ one set of new horns and anvils and one worn set to simulate different tool conditions. The system should possess the following features: (1) Affordability and independence from external computing resources, such as remote servers, to facilitate large-scale deployment at a reasonable cost. (2) High performance, particularly a high accuracy. (3) Robustness against potential concept drift, ensuring that the system can be trained with a limited amount of data covering incomplete patterns while maintaining high accuracy with unseen data.

In our system, the factor that poses a challenge to its generalizability is the change in surface conditions, where we simulate a contaminated surface by dropping cutting fluid on the welding spot. As discussed in the previous subsection, numerous factors may vary across different UWM application scenarios, leading to changes in sensor reading patterns. We opt to alter the workpiece surface condition while keeping all other factors constant (material and dimensions of workpieces, UWM model, and welding process parameters), as it presents the greatest challenge. The introduction of cutting fluid can cause the workpiece to slip during welding, leading to more unpredictable and varied pattern changes. Furthermore, simulating contaminated surface conditions is both practical and reflective of real-world production scenarios. In contrast, factors such as the change of welding amplitude result in predictable changes (amplitude increase in certain frequency ranges) in sensor readings and are typically consistent within a single assembly line, making them less intriguing and challenging compared to surface condition changes. Based on these assumptions, a trained system can be reused on other UWMs with identical models and welding settings, a common configuration for UWMs within the same assembly line.

\section{WeldMon Design}
\label{sec:system_design}

\begin{figure}
  	\centering
  	\includegraphics[width=1\linewidth]{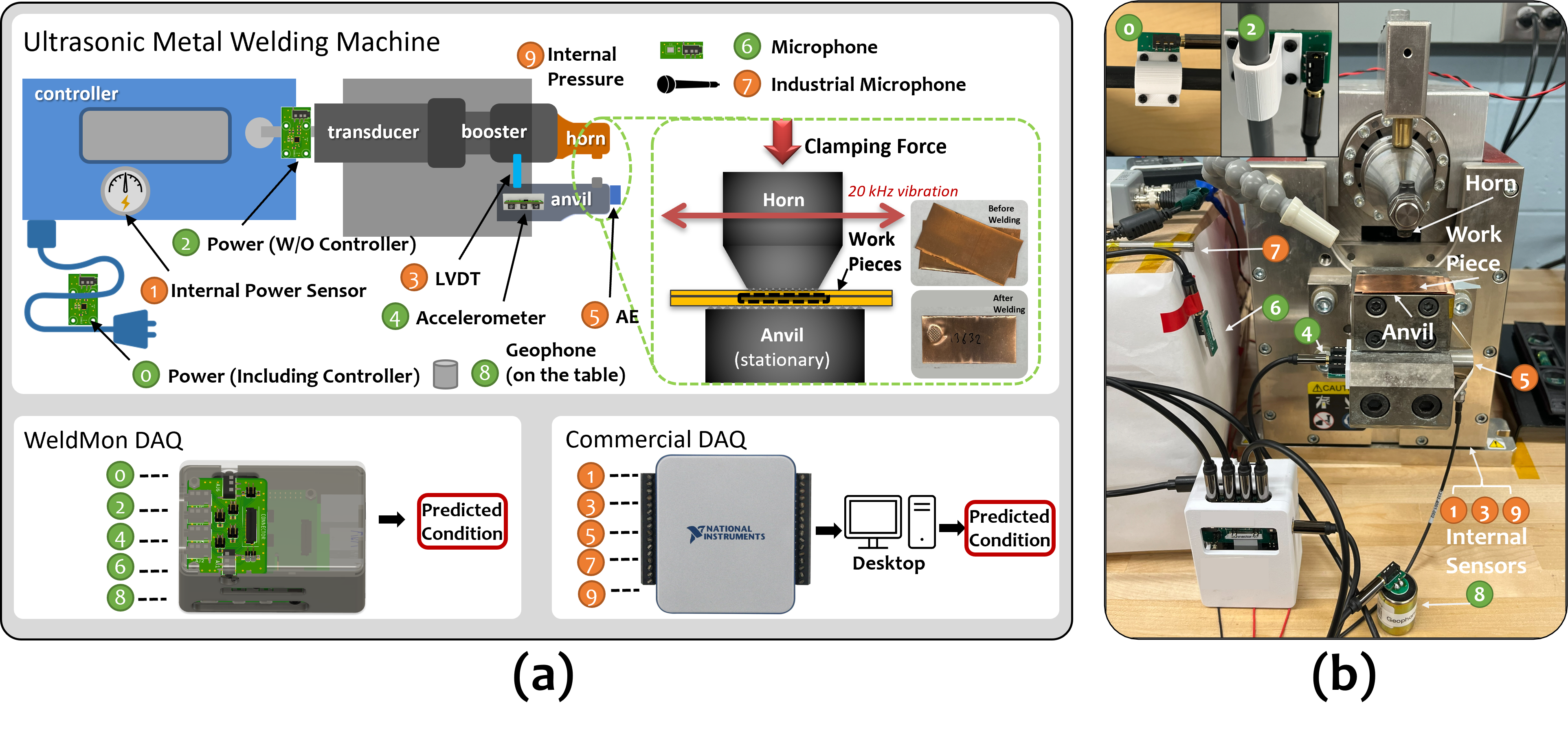}
  	\vspace{-8mm}
	\caption{\textbf{(a)} Overview of the architecture of the UWM, WeldMon, and a widely used commercial system in previous works. \textbf{(b)} Experiment setup of the two systems on a UWM located in a scientific lab. }
        \vspace{-5mm}
	\label{fig:weldmon_architecture}
\end{figure}

\subsection{System Overview}
Fig. \ref{fig:weldmon_architecture}\textbf{(a)} illustrates the architecture of WeldMon and a commercial system, both deployed on the same UWM for performance comparison. In this subsection, we provide an overview of these two systems.

\noindent \textbf{WeldMon:}
WeldMon is a specialized UWM condition monitoring system, featuring a customized DAQ system and a software suite designed for reliable data collection and analysis. The DAQ system comprises five analog sensors, strategically positioned on or near the UWM for optimal performance. These sensors connect to the DAQ device using audio cables, allowing the device to continuously sample the outputs from all analog sensors via an audio ADC. The resulting data is stored in digital formats, and once a predetermined amount of data is collected, a local data analysis algorithm processes the information to estimate the UWM's current condition.

\noindent \textbf{Commercial System:}
The commercial system, similar to that described in \cite{nazir2021online}, incorporates five distinct sensor types, a commercial-grade DAQ device, and a nearby desktop computer for operating data collection and analysis software. The sensors connect to the DAQ device using cables. The collected data is then transferred and stored on the desktop computer via a cable for further processing. Upon completion of data collection, offline data analysis is conducted to classify the UWM's conditions.

\subsection{Customized DAQ System}

\noindent\textbf{DAQ Device:}
Traditional DAQ devices sample analog sensors' outputs using general ADCs. In our system, we opt for an audio ADC instead of a general ADC to minimize costs. Audio ADCs are more cost-effective than general ADCs with the same number of channels, sampling frequency, and sample width (number of bits per sample). Despite its cost advantage, the audio ADC has a few drawbacks: (1) poor frequency response for data below 1000 Hz due to the internal high-pass filter, rendering it unsuitable for measuring slowly changing physical quantities; (2) higher noise levels compared to general ADCs; (3) lack of accurate absolute readings, making it inappropriate for tasks requiring precise sensor readings. These limitations may pose significant challenges for many sensing tasks. However, they do not negatively impact our application scenario for the following reasons: (1) all physical quantities measured in our system exhibit high-frequency characteristics; (2) the additional noise introduced by the audio ADC is acceptable for our specific application; (3) our algorithm does not rely on absolute sensor readings. Instead, we normalize our data, which has proven to yield better results.

\noindent\textbf{Sensors:}
In many previous studies, researchers have preferred using industrial-grade sensors to ensure high-quality data. However, the high cost of these sensors can be a barrier to large-scale deployment, particularly in budget-constrained scenarios. With advancements in sensor technologies, a variety of cost-effective sensors have become available on the market. In our system, we opt for these sensors and strive to achieve sensing performance comparable to that of industrial sensors.

The commercial system measures various physical quantities like sound, anvil vibrations, and UWM current drain, which reliably indicate UWM tool conditions. We select similar sensors for these measurements, excluding LVDT and pressure sensors, as they require internal UWM access and are impractical for our system.

To capture sound signals, an analog-output MEMS (Micro-electromechanical systems) microphone with a wide frequency response is sufficient, as it can adequately capture the required 20kHz sound signal.

Magnetic field-based current sensors can estimate the relative current used by a UWM without breaking its internal circuit, unlike traditional current sensors. However, conventional magnetic field-based sensors need to clamp onto a single conducting wire, requiring opening the UWM's NMB cable, which is impractical. The magnetic fields from the hot and neutral wires in the NMB cable cancel each other out, making the resulting field too small for typical sensors. Flux-gate-based magnetic field sensors are sensitive enough to detect this small magnetic field, which remains approximately proportional to the current drained by the UWM.

For measuring anvil vibrations, a 3-axis MEMS accelerometer with an analog output can replace the acoustic emission sensor. We remove all output filter capacitors that were used in the recommended circuit design. Our experiments show that the sensor, when used in this way, responds well to 20kHz signals, surpassing the minimum requirement for UWM condition monitoring. Additionally, we opt for a sensor that can measure vibration without direct contact with the UWM\footnote{Microhone satisfy this requirement but may have a privacy issue.}, as there may be cases in production where attaching sensors to the UWM is not allowed. We choose a geophone that can be placed on the workbench under the UWM to capture vibrations from the UWM as they propagate through the workbench.

\begin{figure}
  	\centering
  	\includegraphics[width=1\linewidth]{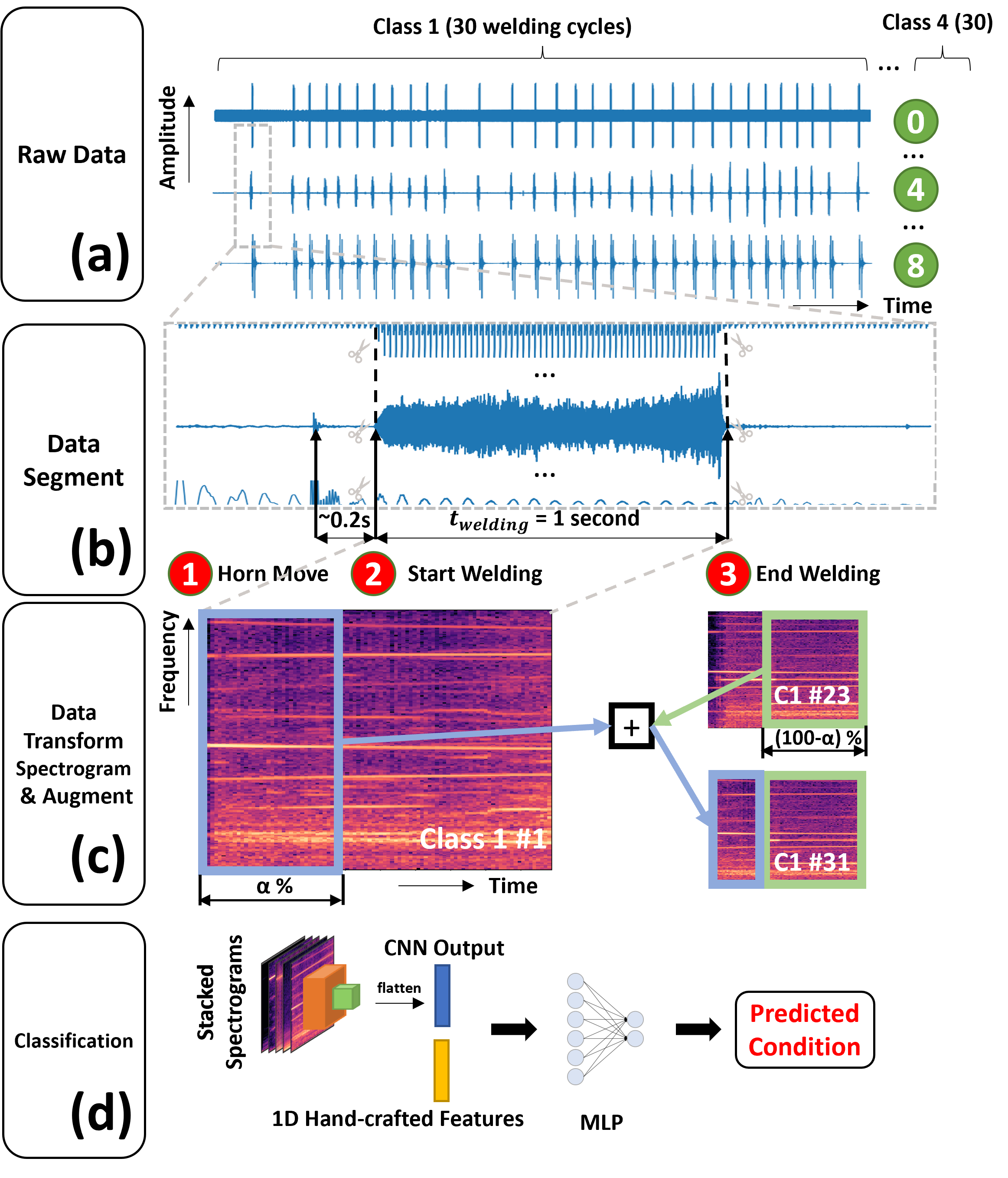}
  	\vspace{-5mm}
	\caption{Weldmon Data Analysis Pipeline.}
        \vspace{-5mm}
	\label{fig:data_analysis_pipeline}
\end{figure}

\subsection{Data Processing Pipeline}
\label{subsec:data_processing_pipeline}
Fig. \ref{fig:data_analysis_pipeline} illustrates the data analysis pipeline used in WeldMon, from the raw data input to the classification result output. In the following subsections, we explain the major modules and various design choices made to enhance data analysis performance.

\noindent\textbf{Raw Input Data: }
The raw input data consists of continuous sensory streams, as illustrated in Fig. \ref{fig:data_analysis_pipeline}\textbf{(a)}. Each second, we obtain $k \times f$ samples, where $k$ represents the number of connected sensors and $f$ denotes the sampling rate of the DAQ device. Instead of using every time window, we only perform data classification on the sensory data collected during the actual welding phase, due to the relevance of this data duration to the tool condition. Qualified data segments are extracted using a data segmentation algorithm.

\noindent\textbf{Data segmentation: }
The data segmentation algorithm aims to extract data collected during the actual welding phase. A toy example is shown in Fig. \ref{fig:data_analysis_pipeline}\textbf{(b)}, where the sensory data in the middle is collected from the accelerometer. Our algorithm consists of two stages. In the first stage, we apply a threshold to identify the start time of Horn Move \redcircled{1}, denoted as $t_{hm}$. Based on prior knowledge, the horn moves for a maximum of 0.2 seconds. In the second stage, we use another threshold to find the timestamp of Start Welding \redcircled{2} after $t_{hm} + 0.2$, and the result is denoted as $t_{sw}$. Since the timestamp obtained from a threshold-based method will be delayed\footnote{The delay originates from the nonoptimal threshold choice.}, we subtract 0.01 second from $t_{sw}$ for a slight adjustment. The End Welding \redcircled{3} timestamp, denoted as $t_{ew}$, is $t_{sw} + t_{welding}$, where $t_{welding}$ is a preset UWM parameter, which is set to 1 second in our case. $t_{sw}$ and $t_{ew}$ are then utilized to extract data segments for all sensory streams. The extracted data segment has a shape of $\mathbb{R}^{N \times k}$, where $N = f \times t_{welding}$ represents the number of samples for each sensor in a data segment.

\noindent\textbf{Data Transformation: }
Since analyzing time-series data in the time-frequency domain has demonstrated potential for better performance compared to the original time domain, we also transform each sensor's data segment into the time-frequency domain using digital Short Time Fourier Transform (STFT) with proper parameters. We only utilize the magnitude of the STFT result. The outcome is commonly known as a spectrogram, which offers a visual depiction of how the frequency spectrum of the input signal changes over time. We further convert the result to decibel units to reduce the range and reveal more details, using the maximum value of the sensor's data segment as a reference. The final transformation result for each sensor has a shape of  $\mathbb{R}^{T \times F}$, where T denotes the number of time frames and F indicates the number of frequency bins. An example is displayed in Fig. \ref{fig:data_analysis_pipeline}\textbf{(c)}, where the brightness of the color is directly proportional to the power of a specific frequency at a given time. There is a difference when we transform the data collected by the commercial system since it spans a larger frequency range (to 100 kHz). We use Mel-spectrograms instead of spectrograms since we want to compress the high-frequency range and mainly use the data range around 20 kHz.

\noindent\textbf{Data Augmentation: }
Data augmentation can enhance data analysis performance when there is insufficient training data to cover the potential feature space. Furthermore, it can help make the trained model more adaptable to unseen data, thereby alleviating the concept drift problem. Despite its many benefits, generating unseen yet realistic samples is a challenge when applying data augmentation. Various data augmentation techniques, such as time warping, adding noise, and time and frequency masking, have been extensively explored in previous works \cite{park2019specaugment}, but they may not be optimal for our data. Given the characteristics of the collected data, we found an effective strategy by combining two spectrograms of the same class using a random split ratio to generate a new spectrogram, as described in Algorithm \ref{alg:data_augmentation}. In the algorithm, $A^{(b)}_{c}$ represents the $c^{th}$ sample of training data $A$ with label $b$. The augmentation factor controls the quantity of augmented training data by multiplying it with the original training data count. For example, with 100 training samples and an augmentation factor of 5, we will get $(5 \times 100) + 100 = 600$ training samples after data augmentation. A toy example is shown in Fig. \ref{fig:data_analysis_pipeline}\textbf{(c)}. In this example, we randomly select two spectrograms (\text{\#}1 and \text{\#}23) in class 1 and a random split ratio $\alpha$. The new spectrogram (\text{\#}31) of class 1 is the concatenation of the first $\alpha\%$ of \#1 and the last $(100-\alpha)\%$ of \text{\#}23.

\begin{algorithm}[t]
\caption{Data Augmentation}\label{alg:data_augmentation}
\begin{algorithmic}[1]
\State \textbf{Input: } $X$, $y$, $\text{aug\_factor}$, $L$ \Comment{$X$, $y$ are input list and label list. $L$ = Set of all labels}
\State $X_{aug} = X, y_{aug} = y$ \Comment{Augmented results}
\For{$l$ in $L$}
    \State $X^{(l)}, y^{(l)}$ \Comment{Train data with label equal to $l$}

    \State $N_{aug} = \text{aug\_factor} \times \text{len}(y^{(l)})$

    \For{$n$ in \text{Range(}$N_{aug}$\text{)}}    
        \State $\alpha \gets \text{rand}(0, 1)$
        \State $p, q \gets \text{randInt}[1, \text{len}(y^{(l)})]$ \Comment{No replacement}
        \State $X_{new\_1} = X^{(l)}_{p}[:\lfloor \alpha \times \text{len}(X^{(l)}_{p}) \rfloor)$ 
        \State $X_{new\_2} = X^{(l)}_{q}[\lfloor \alpha \times \text{len}(X^{(l)}_{q}) \rfloor:]$ 
        \State $X_{new} = \text{concatenate}(X_{new\_1}, X_{new\_2})$
        \State $X_{aug}^{(l)} = \text{append}(X_{aug}^{(l)}, X_{new})$
        \State $y_{aug}^{(l)} = \text{append}( y_{aug}^{(l)}, l)$
    \EndFor
\EndFor
\State \Return $X_{aug}$, $y_{aug}$
\end{algorithmic}
\end{algorithm}

\noindent\textbf{Classification: }
Spectrograms derived from chosen sensors can be stacked to form a multi-channel image, as illustrated in Fig. \ref{fig:data_analysis_pipeline}\textbf{(d)}. These images are used as inputs for a Convolutional Neural Network (CNN), an architecture selected for its demonstrated success in image classification tasks. The output of the last convolution layer is flattened to create an auto-generated feature vector, which is then concatenated with a hand-crafted feature vector. This combined vector is fed into a Multilayer Perceptron (MLP) to predict the condition. We name this method a hybrid method. We employ the same set of hand-crafted features as described in \cite{meng2022physics}, owing to their reported strong performance. Apart from CNNs, we also experimented with Convolutional Recurrent Neural Networks (CRNNs), which are anticipated to perform better when capitalizing on the temporal information present in spectrograms. However, in our dataset, CRNNs were not as effective, likely due to the limited temporal information available in the spectrograms.

\section{Implementation}
\label{sec:implementation}

In this subsection, we describe the hardware and software implementation details for WeldMon and the commercial system, as well as the data collection process. We open-source our hardware and software implementation, and all resources will be found on GitHub at \url{https://github.com/beitong95/WeldMon_Public.git}.

\begin{table}[]
\caption{Hardware Implementation Details}
\label{tab:hardware_implementation}
\begin{tabular}{lllr}
\hline
\textbf{System}                      & \textbf{Device}         & \textbf{Model}                                                                            & \textbf{Price (\$)}        \\ \hline
\multirow{6}{*}{WeldMon}    & \greencircled{0} \& \greencircled{2} Power             & DRV425                                                                           & 10                \\
                            & \greencircled{4} Accelerometer     & ADXL335                                                                          & 15                \\
                            & \greencircled{6} Microphone        & ICS-40300                                                                        & 5                 \\
                            & \greencircled{8} Geophone          & SM-24                                                                            & 60                \\
                            & DAQ Device        & \begin{tabular}[c]{@{}l@{}}R-Pi4 + Respeaker \\ Voice HAT\end{tabular} & 94                \\
                            & Total             &                                                                                  & \textless 200     \\ \hline
\multirow{7}{*}{Commercial} & \orangecircled{1} Power             & Internal                                                                         &                   \\
                            & \orangecircled{3} LVDT              & Internal                                                                         &                   \\
                            & \orangecircled{5} Acoustic Emission & R15$\alpha$                                                         & 500               \\
                            & \orangecircled{7} Microphone        & GRAS 40 PP                                                                       & 700               \\
                            & \orangecircled{9} Pressure        & Internal                                                                       &                \\
                            & DAQ Device        & NI USB 6361                                                                  & 2561              \\
                            & Total             &                                                                                  & \textgreater 3500 \\ \hline
\end{tabular}
\vspace{-5mm}
\end{table}
\noindent\textbf{WeldMon Hardware:}
The assembled system is depicted in Fig. \ref{fig:weldmon_architecture}\textbf{(b)}. Each sensor is utilized with a custom printed circuit board (PCB), designed using Eagle software and manufactured by JLCPCB. The edge device consists of a Raspberry Pi 4, a Respeaker audio accessory HAT, and a customized connector board. The 3D-printed enclosure for power sensors and the DAQ device, and the sensor mounting base for the accelerometer are designed with Fusion 360 and printed using a Creality Ender 3 3D printer. Shielded 3.5 mm audio cables are employed to connect sensors to the edge device, and data is sampled at a 48 kHz rate with 16-bit sample width. WeldMon utilizes a power outlet as its energy source, which is readily available in most welding workspaces. Additional details, such as sensor type, model, and price, can be found in Table \ref{tab:hardware_implementation}.

\noindent\textbf{Commercial Hardware:}
The data acquisition (DAQ) device used is from National Instruments. The power sensor, pressure sensor, and LVDT sensor are internal sensors, and their data is read by the NI DAQ through a customized interface on the UWM. Sound signals are collected by a GRAS 40 PP microphone, powered and amplified by a GRAS 12AL. The acoustic emission sensor, R15$\alpha$, uses a preamplifier to amplify the signal so it falls within NI DAQ's measurement range. Sensors are connected to the NI DAQ using Dupont cables and are sampled at 200 kHz.

\noindent\textbf{UWM and Workpiece Parameters:}
In our experiment, we employed a Branson Ultraweld L20 model as the ultrasonic welding machine (UWM). We set the operating parameters as follows: pressure at 50 Psi, amplitude at 40 µm, and welding duration at 1 second. The copper workpieces used in the study measured 1 inch in width, 2 inches in length, and had a thickness of 0.008 inches.

\noindent\textbf{Software and Training Setup:}
The WeldMon data collection software module is developed using Python. Meanwhile, the data collection software for NI DAQ is programmed in MATLAB. We implemented data processing pipelines for both systems in Python. 

Our workstation for model training is equipped with an Intel Core i9-10900X CPU @ 3.70GHz, 64 GB RAM, and an Nvidia RTX3080 GPU. We employ the basic CNN architecture without any fine-tuning to the neural network's structure, which consists of four convolution layers, and three fully connected layers (roughly 11 million parameters). We experiment with popular CNN architectures, such as ResNet-50, but observed inferior results compared to using basic CNN architecture. The Adam optimizer with a 4e-4 learning rate is used for training, with early stopping based on training accuracy and a learning rate reduction strategy. We utilize a batch size of 32 and a maximum of 100 epochs, with the typical training time for each model being under one minute. Evaluation takes place on the same workstation, except for algorithm latency testing, which is performed on the Raspberry Pi 4.

\begin{table}[]
\caption{Tool and Surface Condition Explanation}
\label{tab:condition_explanation}
\begin{tabular}{ll}
\hline
\textbf{Condition}    & \textbf{Explanation}                          \\ \hline
New          & The horn and anvil are new ($<$ 1000 usage times).       \\
Worn         & The horn and anvil are worn ($>$ 1000s usage times). \\
Clean        & The workpiece surface is clean.                           \\
Contaminated & The workpiece has cutting fluid drops on its surface.           \\ \hline
\end{tabular}
\vspace{-5mm}

\end{table}

\noindent\textbf{Data Collection:}
In our study, we focus on two tool conditions and two workpiece surface conditions, as described in Table \ref{tab:condition_explanation}. We install WeldMon and the commercial system on the same UWM and perform 30 welding cycles for each combination of tool and surface conditions (New + Clean, New + Contaminated, Worn + Clean, and Worn + Contaminated). This procedure yields two datasets for each system containing sensory data from the same 120 welding cycles across four distinct classes. The data are then processed using the data processing pipeline, as detailed in Section \ref{subsec:data_processing_pipeline}. The four classes can be combined for various classification tasks. For instance, if the objective is to classify the tool condition, we can merge the data from New + Clean and New + Contaminated classes and assign a new class label, "New," which would be the same for data in the worn condition.
\section{Evaluation}
\label{sec:evaluation}

\begin{figure*}[t]
    \centering
    \includegraphics[width=1\linewidth]{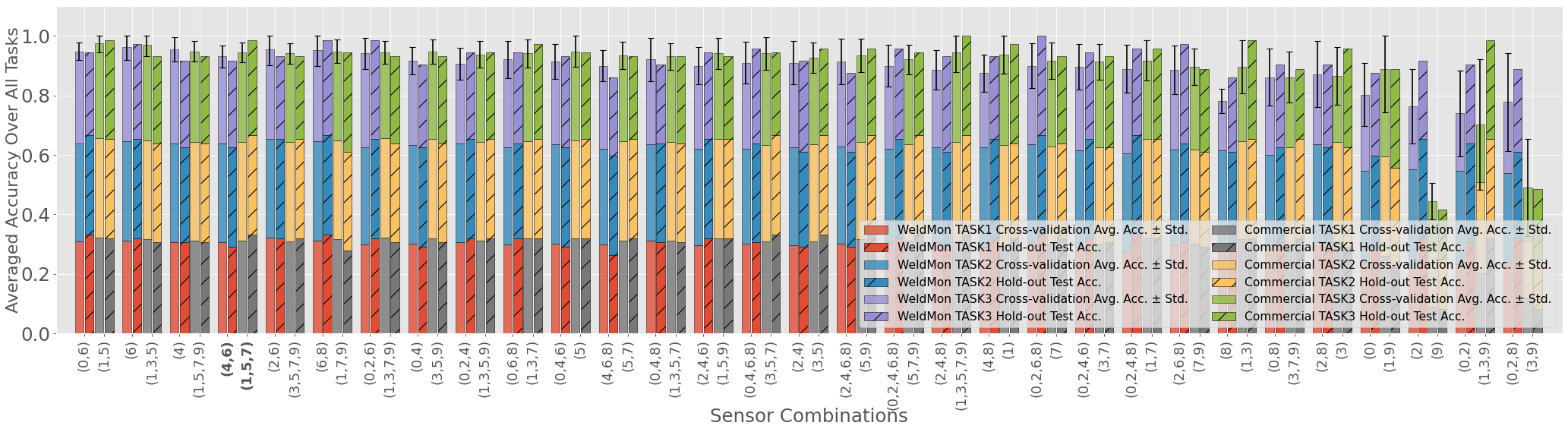}
    \caption{Average Classification Accuracy Across All Tasks for Different Sensor Combinations. The x-axis shows the combination of sensor IDs (\textbf{bolded} ones indicate the selected sensor combinations), which are the same as those used in Fig. \ref{fig:weldmon_architecture}.}
    \label{fig:compare_sensor_combination}
     \vspace{-5mm}

\end{figure*}

\noindent In the upcoming evaluation section, we aim to address the following key questions:

\begin{itemize}[leftmargin=*]
\item What are the optimal sensor combinations for UWM condition monitoring in each system?
\item How do data augmentation and the volume of augmented data impact classification accuracy?
\item Does our classification method outperform existing approaches?
\item Is WeldMon a viable replacement for Commercial system?
\end{itemize}

\noindent\textbf{Task Setup:}
We design three classification tasks based on the collected data to comprehensively evaluate the effectiveness of our system. Task 1 is a multiclass classification of mixed tool and surface conditions. Task 2 and Task 3, on the other hand, focus solely on binary classification of tool conditions, ignoring the surface condition. Task 2 is trained using data from all mixed conditions, while Task 3 is trained only with New + Clean and Worn + Clean data. To simulate the common concept drift problem in real-world scenarios, we also test Task 3 with data collected from the Contaminated condition. Task 1 presents a challenge due to its multiclass classification nature, while Task 3 is even more challenging as it requires the model to accurately classify unseen data. 
 

\noindent\textbf{Evaluation Process \& Metrics:}
For each task and parameter setting (sensor combination, augmentation factor, classification method, etc.), we begin by dividing the dataset into a training set and a held-out testing set, using an 80:20 ratio\footnote{We use the same random state when splitting the data to make sure the training and testing datasets are the same when we change other parameters. }. Next, we perform stratified repeated cross-validation (CV) on the training set. During this process, we augment the training folds according to a predetermined augmentation factor before training each fold. Following cross-validation, we train a model using the entire training set (also augmented based on the augmentation factor) and evaluate it on the held-out testing set. This yields two sets of accuracy measurements: the average cross-validation accuracy, accompanied by its standard deviation, and the final testing accuracy. While the average cross-validation accuracy and its standard deviation are sufficient for comparing various model designs, we also report the testing accuracy to provide a more comprehensive and robust evaluation of the model's performance.

\noindent\textbf{Baseline Methods:}
To assess the effectiveness of our proposed classification method, we compare our method with a state-of-the-art method for UWM condition monitoring\cite{meng2022physics}. This method employs DWT (Discrete Wavelet Transform) to extract multiple features from the data. A multi-layer fully connected neural network is then used to predict the condition based on selected features. We implement the baseline method following the algorithm described in \cite{meng2022physics}. We also implement a CNN-only method where we remove the handcrafted features used in Fig. \ref{fig:data_analysis_pipeline}\textbf{(d)}.

Task 1 can also be solved by ensemble learning, combining outputs of two binary classifiers, but the result is not as good as using a multiclass classifier based on our experiment results. 




\subsection{Identifying Optimal Sensor Combinations}
\label{subsec:choose_sensor_combination}
For each system, we assess all possible sensor combinations (31 in total) and their corresponding accuracies. For each combination, we train and test our hybrid classification model using data from the selected sensors to determine accuracy. The augmentation factor is set to 0, and cross-validation repetition is 1. We rank sensor combinations based on the cross-validation LCB\footnote{Calculated by cross-validation accuracy mean - 1.96 standard deviations} (Lower Confidence Bound), as shown in Fig. \ref{fig:compare_sensor_combination}. It is evident that the accuracies of both systems are sensitive to changes in sensor combinations, highlighting the significance of identifying optimal sensor configurations for improved performance. We also rank the sensor combinations for the other two baseline methods to find sensor combinations that perform well in all three classification methods\footnote{Due to page limitations, the other two images can be found in our GitHub repository.}. For the WeldMon system, we select Accelerometer \greencircled{4} and Microphone \greencircled{6} as the best sensor combination, as it ranks among the top 4 for all classification methods. For the Commercial system, the optimal sensor combination includes Power \orangecircled{1}, Acoustic Emission \orangecircled{5}, and Microphone \orangecircled{7}, as it stably ranks within the top 5 for all classification methods. These best sensor combinations will be used for subsequent evaluations.

Several intriguing observations can be made from the Fig. \ref{fig:compare_sensor_combination}. Using a single sensor in WeldMon, such as an accelerometer or a microphone, yields excellent results, suggesting the potential for removing some sensors to reduce costs and save space in WeldMon (below \$120). It can be seen that adding more sensors or replacing existing ones may not significantly enhance performance, as demonstrated by combinations (4) and (4, 6), since the new sensory stream might be correlated with the original one. Using incorrect sensors only can drastically reduce performance, such as Geophone and Power sensors in WeldMon and LVDT and Pressure sensor in the Commercial System, as they all show bad accuracy in Fig. \ref{fig:compare_sensor_combination}. Fortunately, combining poor sensors with good ones does not considerably diminish performance, as demonstrated by combinations (0, 2, 4, 6, 8) and (0, 2, 8).

\begin{figure}[t]
  	\centering
  	\includegraphics[width=1\linewidth]{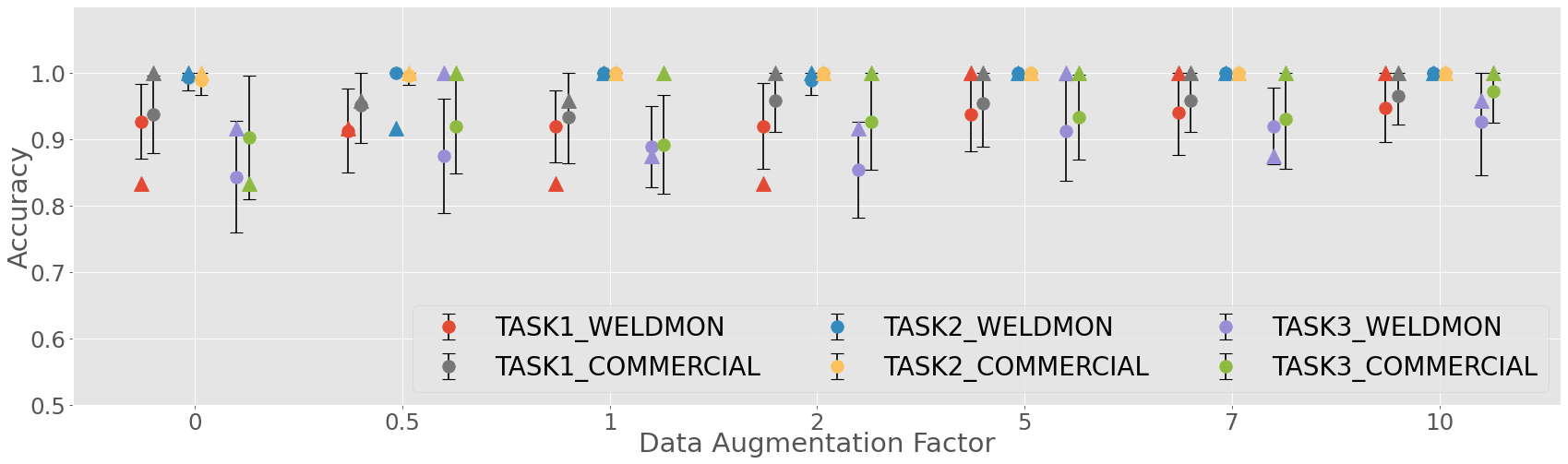}
        \vspace{-5mm}
	\caption{Error-bar plot of Classification Accuracy under Different Data Augmentation Factors. The dots show the average cross-validation accuracy. The Upward-pointing triangles show the accuracy tested on the held-out data. }
	\label{fig:compare_augmentation_factor}
        \vspace{-5mm}
\end{figure}

\subsection{Effectiveness of Data Augmentation}
\label{subsec:effectiveness_of_data_augmentation}
We test our system under different data augmentation factors ranging from 0 to 10 and report the averaged accuracy and standard deviation of 3 repetitions of 6-fold cross-validation and also the final testing accuracy as shown in Fig. \ref{fig:compare_augmentation_factor}. We can find a clear trend that the results of all tasks of both systems improve and get more reliable, and the accuracy gap between WeldMon and the commercial system gets smaller with the increment of the augmentation factor. When the augmentation factor reaches 5, both systems reach their optimal performance and have the smallest final test accuracy differences. More augmentation will not help to improve the performance significantly. Recall in Task 3, both systems are challenged with the concept drift problem. With the data augmentation factor set as 10, the averaged accuracy of cross-validation improves by 8.3 \% and 6.9 \% for WeldMon and Commercial system, respectively which proves the effectiveness of the proposed data augmentation method in alleviating the concept drift problem. For Task 1, the final test accuracy improves by 16.7 \% to 100 \% for WeldMon with the augmentation factor set to 5 or more which shows the great impact of the proposed data augmentation method towards the classification accuracy of WeldMon. 

\begin{table}[t]
\vspace{-5mm}

\centering
\begin{threeparttable}
\caption{Accuracy \& Latency Under Different Methods}
\begin{tabular}{cccccc}
\hline
System                                                                       & Task                   & Method  & CV\tnote{1}            & Test  & Time\tnote{2} (ms) \\ \hline
\multirow{9}{*}{\begin{tabular}[c]{@{}l@{}}WeldMon\\ 48kHz\end{tabular}}     & \multirow{3}{*}{Task1} & Hybrid & 0.941 (0.057) & 0.958 & 383 (764)     \\
                                                                             &                        & DWT    & 0.910 (0.079) & 0.917 & 265 (607)     \\
                                                                             &                        & CNN    & 0.938 (0.059) & 1.000 & 266 (627)      \\ \cline{2-6} 
                                                                             & \multirow{3}{*}{Task2} & Hybrid & 0.997 (0.014) & 1.000 & 386 (764)     \\
                                                                             &                        & DWT    & 0.983 (0.035) & 1.000 & 263 (681)     \\
                                                                             &                        & CNN    & 1.000 (0.000) & 1.000 & 265 (682)      \\ \cline{2-6} 
                                                                             & \multirow{3}{*}{Task3} & Hybrid & 0.906 (0.063) & 0.875 & 384 (764)     \\
                                                                             &                        & DWT    & 0.920 (0.072) & 1.000 & 262 (612)     \\
                                                                             &                        & CNN    & 0.913 (0.073) & 0.917 & 263 (649)      \\ \hline
\multirow{9}{*}{\begin{tabular}[c]{@{}l@{}}Commercial\\ 200kHz\end{tabular}} & \multirow{3}{*}{Task1} & Hybrid & 0.955 (0.072) & 0.958 &      \\
                                                                             &                        & DWT    & 0.899 (0.076) & 0.917 &      \\
                                                                             &                        & CNN    & 0.951 (0.064) & 1.000 &      \\ \cline{2-6} 
                                                                             & \multirow{3}{*}{Task2} & Hybrid & 1.000 (0.000) & 1.000 &      \\
                                                                             &                        & DWT    & 0.986 (0.033) & 1.000 &      \\
                                                                             &                        & CNN    & 1.000 (0.000) & 1.000 &      \\ \cline{2-6} 
                                                                             & \multirow{3}{*}{Task3} & Hybrid & 0.951 (0.074) & 1.000 &      \\
                                                                             &                        & DWT    & 0.851 (0.066) & 0.833 &      \\
                                                                             &                        & CNN    & 1.000 (0.000) & 1.000 &      \\ \hline
\end{tabular}
\begin{tablenotes}
  \footnotesize
  \item[1] Accuracy of 3 repetitions of 6-fold cross-validation, with results reported as mean (standard deviation). 
  \item[2] Mean (Max) computation time per sample over 300 experiments in ms.
\end{tablenotes}
\label{tab:compare_data_analysis_methods}
\vspace{-5mm}
\end{threeparttable}
\end{table}

\subsection{Comparing Different Classification Methods} 
\label{subsec:compare_methods}
In this experiment, we set the augmentation factor to 5 and employ the optimal sensor combination for each system. We then compare the accuracy of our hybrid method and two baseline methods under different systems and tasks. As shown in Table \ref{tab:compare_data_analysis_methods}, our method shows similar performance on both systems compared to the CNN-only method and outperforms the DWT method in most cases. During the experiment, we also find the performance of our method is less sensitive to different sensor combinations compared to the CNN-only method. Also, our method is easier to converge but the CNN-only method may not converge in some cases.  

To compare the computation time of the proposed and baseline methods, we run 300 data processing cycles (from data transformation to the end of classification) on a Raspberry Pi 4 with 4 GB RAM, while simultaneously collecting data. For each task and model combination, we report the mean and maximum computation time in Table \ref{tab:compare_data_analysis_methods}. Results show that the proposed hybrid method has longer computation time compared to the other two baseline methods, due to extra feature extractions and more complex neural network model. The observed latencies are acceptable, considering that the minimum interval between two welding processes is much longer than the average computation time. In future work, we aim to further reduce data processing time, enabling our algorithm to run on devices with more limited computing resources and handle additional sensory streams or more complex algorithms.

\begin{figure}[t]
    \centering
    \includegraphics[width=1\linewidth]{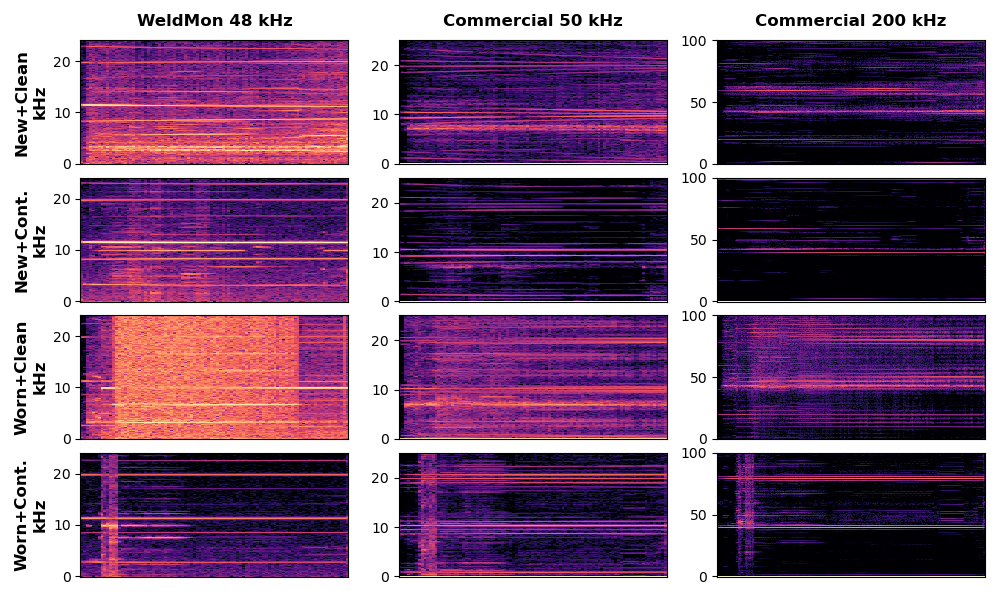}
    \caption{Sample spectrograms of WeldMon (accelerometer), Commercial System (acoustic emission sensor), and Commercial System with downsampled data under various mixed conditions. Each row corresponds to a specific mixed condition, while each column indicates the system name and its corresponding sampling rate. The x-axis spans 1 second.\protect\footnotemark}
    \label{fig:all_spectrograms}
    \vspace{-5mm}
\end{figure}
\footnotetext{Cont. is short for contaminated. For a more accurate comparison, linear spectrograms are used for the commercial system, rather than Mel-spectrograms which are used in other evaluations.}

\subsection{Comparing WeldMon with the Commercial System}
\label{subsec:compare_systems}

\begin{table}[t]
\centering

\begin{threeparttable}

\caption{Accuracy of Different Systems}
\begin{tabular}{lcccccc}

\hline
\multirow{2}{*}{\begin{tabular}[c]{@{}l@{}}System\\ Sampling Rate\end{tabular}} & \multicolumn{2}{c}{Task1} & \multicolumn{2}{c}{Task2} & \multicolumn{2}{c}{Task3} \\ \cline{2-7} 
                                                                                & CV\tnote{1}          & Test        & CV          & Test        & CV          & Test        \\ \hline
WeldMon 48kHz                                                                   & 0.94        & 0.96        & 1.00        & 1.00        & 0.91        & 1.00\tnote{2}        \\
Commercial 200kHz                                                               & 0.94        & 0.96        & 1.00        & 1.00        & 0.92        & 1.00        \\ 
Commercial 50kHz                                                                & 0.97        & 1.00        & 1.00        & 1.00        & 0.96        & 1.00        \\ \hline
\label{tab:compare_systems}
\end{tabular}
\begin{tablenotes}
  \footnotesize
  \item[1] Averaged accuracy of 3 repetitions of 6-fold cross-validation. 
  \item[2] The test accuracy is different from Table \ref{tab:compare_data_analysis_methods} may be caused by randomness in data augmentation. 

\end{tablenotes}
\end{threeparttable}
\vspace{-5mm}
\end{table}

Despite using optimal configurations, a significant accuracy disparity persists between WeldMon and the commercial system in Task 3, as shown in Fig. \ref{fig:compare_augmentation_factor} and Table \ref{tab:compare_data_analysis_methods}. Our goal is to investigate the root cause of this discrepancy. We begin by visually comparing spectrograms from different systems. Fig. \ref{fig:all_spectrograms} presents sample spectrograms for accelerometer and acoustic emission sensor data, as well as downsampled acoustic emission sensor data at 50 kHz. The commercial system's spectrograms display unique information above 24 kHz, which may contribute to the accuracy difference. Comparing WeldMon 48 kHz and Commercial 50 kHz, the commercial system's spectrograms exhibit more detail and less noise, suggesting sensor specification differences could contribute to the accuracy gap.

We then conduct a quantitative analysis using a specially designed experiment. To ensure a meaningful comparison, both systems must use identical sensor types. We select the accelerometer and acoustic emission sensor, which monitor anvil vibrations, and a pair of microphones for each system. We also downsample the commercial system's data to 50 kHz to evaluate the impact of the maximum frequency limit. We apply our data analysis pipeline to the data from the chosen sensors (with an augmentation factor of 5) and present the findings in Table \ref{tab:compare_systems}. The results reveal a substantially reduced accuracy gap between WeldMon and the commercial system sampling at 200 kHz, emphasizing the importance of the additional power sensor used in Sec. \ref{subsec:compare_methods}. Interestingly, the commercial system sampling at 50 kHz demonstrates superior performance, indicating that the maximum frequency limit is not the main factor behind the accuracy discrepancy. These observations lead us to conclude that the primary source of the accuracy gap lies in the sensors themselves, whether in the sensor type or their specifications. Although sensor differences are challenging to overcome, we can mitigate them with advanced algorithms, sensor fusion techniques, or by adding new sensors, which we intend to explore in our future work.


\section{Related Work}
\label{sec:related_work}
This paper presents a cost-effective welding machine monitoring system that runs on an edge device, delivering real-time, accurate, and reliable results. We review cost-effective data acquisition systems, DNN-based machine condition monitoring, and solutions for handling concept drift.
\subsection{Cost-effective DAQ systems:}
The deployment of machine learning-based monitoring systems relies heavily on the cost of data acquisition devices. Traditional systems used for UWM are expensive, hindering large-scale deployment. Numerous studies have investigated low-cost data acquisition devices for diverse applications. For instance, \cite{gonzalez2018low} developed a cost-effective Arduino-based DAQ system for automotive dynamics, which showed only a 2.19 percent error rate compared to traditional vibration acquisition methods. Despite its success, single-sensor DAQ systems have limitations, as discussed in Sayyad et al. \cite{sayyad2021data}, leading to a preference for multi-sensor approaches. \cite{tian2021senselet++} introduced an affordable multi-sensor DAQ system employing the 1-Wire Bus for large-scale indoor environmental monitoring. However, due to its limited sampling frequency, this system is not suitable for applications such as UWM, which operates at 20 kHz. To address this, Soto-Ocampo et al. \cite{soto2020low} designed an affordable, high-frequency DAQ system, diagnosing a bearing test rig, which is five times cheaper than myDAQ from National Instruments. Another challenge in low-cost data acquisition systems is real-time or near real-time processing. \cite{bedretchuk2023low} presented an IoT-based intelligent low-cost system for vehicle data acquisition, offering near real-time data processing at a lower cost (up to 13 times cheaper) compared to industrial devices. In ultrasonic welding, Lu et al. conducted case studies on reducing the cost of building a UWM condition monitoring system by decreasing the sampling rate and the number of sensors, but they only simulated the result on a high-cost DAQ system and did not implement an actual low-cost system. Moreover, their work relied on hand-crafted features for data processing, which may not be optimal and robust to context changes. Thus, developing a low-cost DAQ system holds significant potential for various applications.

\subsection{DNN-based Machine Condition Monitoring}
Tool Condition Monitoring (TCM) has gathered significant research interest over the past few decades due to its critical role in the manufacturing industry \cite{nazir2021online}. Tool wear mechanisms are influenced by varying process parameters and conditions \cite{li2019data}, complicating the development of effective TCM systems for certain manufacturing processes. Data-driven TCM methods have utilized fuzzy logic systems, Bayesian networks, decision trees, support vector machines (SVM), and artificial neural networks (ANN). Recently, deep learning techniques have emerged as a more effective approach. For instance, Zhao et al. \cite{zhao2017learning} introduced a deep learning model, Convolutional Bi-directional Long Short-Term Memory Networks (CBLSTM), for tool wear testing in CNC milling.


Online DNN-based TCM research for UWM is limited, mainly due to its high oscillation frequency and short welding cycle. In \cite{wu2022end}, the authors employed a convolutional neural network for joint quality prediction in UMW, showcasing robustness to tool conditions. Meanwhile, Lu et al. used a multi-layer perceptron (MLP) classifier for identifying welding disturbances related to tool and material surface conditions. In this paper, we transform multimodal sensory data from welding phases into multi-channel images, then use a CNN for feature extraction and an MLP for classification.

\subsection{Addressing Concept Drift in Monitoring Systems}
Concept drift poses a significant challenge in supervised machine learning-based monitoring, particularly in real-world applications with frequently changing contexts. Various studies have proposed solutions for this issue. For instance, Yang et al. \cite{yang2021cade} developed CADE, a system using a contrastive autoencoder and distance-based explanation method to enhance supervised classifiers in security contexts. In manufacturing condition monitoring, Shi et al. \cite{shi2023reliable} introduced a contrastive generalization net (RFACGN) for intelligent fault diagnosis under unseen machine and operational conditions. This net demonstrates excellent generalization ability and diagnostic efficiency compared to other state-of-the-art methods. Lin et al. \cite{lin2019concept} proposed an ensemble learning algorithm for offline classifiers, addressing concept drifts and imbalanced data in three-stage condition-based maintenance (CBM). For online applications, Zenisek et al. \cite{zenisek2019machine} presented a method to detect concept drift in data streams as an indicator of defective system behavior in industrial settings, validating their approach on synthetic and real-world data sets. These studies highlight the importance of addressing concept drift in machine learning-based monitoring systems, especially in manufacturing tool condition monitoring. We propose a data augmentation method specially used for the UWM monitoring data to alleviate the concept drift problem.
\section{Conclusion}
\label{sec:conclusion}

This paper introduces WeldMon, the first cost-effective autonomous system for real-time UWM condition monitoring. We detail WeldMon's design, implementation, and deployment, emphasizing its cost-effectiveness, superior accuracy, and robustness to context changes. The results indicate that WeldMon offers a practical, effective, and reliable solution for real-world applications, paving the way for future advancements in cost-efficient and robust condition monitoring systems.

\bibliography{refs} 
\bibliographystyle{ieeetr}
\end{document}